\documentclass[%
 reprint,
superscriptaddress,
 amsmath,amssymb,
 aps,
 pra,
]{revtex4-1}

\usepackage[utf8]{inputenc}
\usepackage{xcolor}
\usepackage{graphicx}
\usepackage{amsmath}

\newcommand\w{1.15}

\begin{document}

\title{Syndrome-Derived Error Rates as a Benchmark of Quantum Hardware}
\author{James R. Wootton}
\affiliation{IBM Quantum, IBM Research Europe, Zurich}
\date{\today}

\begin{abstract}

Quantum error correcting codes are designed to pinpoint exactly when and where errors occur in quantum circuits. This feature is the foundation of their primary task: to support fault-tolerant quantum computation. However, this feature could used as the basis of benchmarking: By analyzing the outputs of even small-scale quantum error correction circuits, a detailed picture can be constructed of error processes across a quantum device. Here we perform an example of such an analysis, using the results of small repetition codes to determine the error rate of each qubit while idle during a syndrome measurement. This provides an idea of the errors experienced by the qubits across a device while they are part of the kind of circuit that we expect to be typical in fault-tolerant quantum computers.

\end{abstract}

\maketitle

\section{Introduction and Motivation}

The development of quantum computers requires the benchmarking and validation of quantum hardware~\cite{eisert:2019}. When aiming towards the long-term goal of fault-tolerant and scalable quantum computation, it is crucial for this benchmarking to account for quantum error correction: probing to what degree its requirements are met, and whether it does indeed provide the functionality that it is designed for.

It is with this motivation that many experiments have been performed as proofs-of-principle for the components of quantum error correction circuits as well as demonstrating error suppression in certain restricted cases (see \cite{andersen:20,chen:21,ryananderson:21,hilder:21,gong:21,erhard:21,chen:22} for selected recent examples, and references therein). The most recent examples have made the major breakthrough of demonstrating surface codes of distance $3$~\cite{krinner:2022,zhao:2021}.

In such demonstrations, the metric of success is often defined at a large-scale level. For example, the repetition code implementations of ~\cite{wootton:18,wootton:20} look at the probability of a logical error for increasing code distances. These codes restricted by the fact that only a limited subset of errors can be corrected. However, the code distance for this subset can be made very large even with current quantum hardware. Specifically, for $n$ total qubits used the distance will be $d \approx n/2$. Since the probability of a logical error decays exponentially with $d$, the number of samples required to estimate this probability must therefore increase exponentially. Analysis of progress using only the calculation of this value is therefore not scalable.

It is for this reason that we will turn our attention to the microscopic scale. Quantum error correction codes constantly output syndrome information, which is explicitly designed to detect errors, and pinpoint exactly where and when single error events occur. In the ideal case, we deduce exactly which kind of Pauli error occurred on a specific qubit at a specific point within the circuit. This information is available on a shot-by-shot basis, allowing for the computation to correct or otherwise avoid the effects of the error. Nevertheless, we can also make use of the information when comparing results from many shots. This can be used to compile information about the likelihood of different kinds of error, giving us probabilities for different error events at specific points within a quantum error correction circuit.

These syndrome-derived error rates have already been well-studied as a means to provide accurate weights within a decoding procedure (such as in~\cite{spitz:2018}). However, relatively little attention has been given to their potential for benchmarking (an exception being \cite{wootton:20} and the functionality within Qiskit that it documents~\cite{qiskit}, on which this current work is also built). 

The fact that these error rates are tailored to specific points within a circuit, and that they can only be measured within the context of a large multi-qubit circuit, stands in contrast to many other benchmarking techniques. In randomized benchmarking~\cite{magesan:2011}, for example, results are tailored to specific qubits and specific gate type. Also, the benchmarked gates are implemented in isolation by default. For performance in the context of more complex circuits, we must assume that the same results still hold, or design more sophisticated benchmarks (such as ones designed to be sensitive to cross-talk~\cite{mckay:2020}).

The advantage of the syndrome-derived error rates is that such assumptions or upgrades are not required. From the context of fault-tolerant quantum computation, the most important error rates are those of qubits that are actively part of an error correction circuit. By using the syndrome from such a circuit to analyze the errors within that very circuit, we obtain the most accurate assessment of error rates within that circuit at the microscopic level. Of course, we can still only assume that the same qubits would experience equivalent error rates if used for a different quantum error correcting code. However, this is a much more reasonable assumption than for error rates calculated for a qubit analyzed in isolation.

Different points within a circuit will be subject to different forms of noise. Each error rate will therefore have contributions from multiple different error processes. In this work we will therefore focus on one particular point within a particular circuit: the idling of the central qubit of a $d=3$ repetition code during measurement. This will allow us to add delay gates to increase the idle time, and therefore increase the effects of relaxation noise in particular, to see the effects of $T_1$ and $T_2$ times within the error rates.

\section{Characterizing Code Qubit Idling Errors}

\subsection{A minimal set of experiments}

In a quantum error correcting code, different physical qubits play different roles. Some are the `code qubits' (also known as `data qubits'), whose collective properties are used to store and manipulate quantum information. Others are `auxiliary qubits', used to mediate multi-qubit measurements. Circuits for quantum error correction consist of a constantly repeating schedule of these multi-qubit measurements, which typically involve a set of two-qubit gates between code and auxiliary qubits, followed by measurements of the auxiliaries. This measurement is typically the longest duration gate within circuit. In designing a circuit, it can be advantageous to find a way to keep the code qubits active while some of the auxiliary qubits are being measured. This is to avoid the errors that would occur due to the qubit being idle. Nevertheless, the idle time for a code qubit during measurement of its auxiliaries is a common feature of quantum error correcting circuits. As such, whether we are seeking to avoid them or accepting their presence, it is important to determine the potency of errors on code qubits during measurement of auxiliaries.

To do this we use repetition codes implemented along a line, in the same manner as in~\cite{wootton:20}. A distance $d$ repetition code of this form has $d$ code qubits and $d-1$ auxiliary qubits. There are therefore $2d-1$ qubits in total, alternating along a line between code and auxiliary qubits. Of the code qubits, the $d-2$ in the bulk are part of two distinct syndrome measurements. The auxiliary qubits for these are the two neighbours of each code qubit. The two code qubits on the ends of the line are part of only one syndrome measurement and therefore have less detailed information extracted about their errors. 

Our approach is to select a single qubit on a device whose errors we wish to examine. Since we wish to examine the errors for a code qubit while idle, the selected qubit must obviously be a code qubit of the code we implement. To get sufficiently detailed information, it will also need to be a code qubit within the bulk of the code (not one involved in less syndrome measurements due to being part of the edge). The minimal repetition code to have at least one bulk code qubit is that with $d=3$, where it is the central qubit in a line of five.

When implementing the code, we will run a predetermined number of syndrome measurement rounds. During the first round, errors detected on code qubits could come from a variety of sources, most notably incorrect initialization of the qubit, and imperfections in the two-qubit gates used for the syndrome measurements. The code qubits will subsequently experience their first errors due to delay while the auxiliary qubits are being measured. These will be detected by the second and all subsequent rounds of syndrome measurement, alongside the aforementioned imperfections in the two-qubit gates. During final readout, the code qubits are also measured and the results are used to infer a final effective syndrome measurement round. Results from this round will additionally detect measurement errors for the code qubits. In order to concentrate on errors during the idle time, we therefore require $T \geq 2$ syndrome measurement rounds. Results from the first and (effective) final round will not be used. 

Repetition codes only detect a single type of error, whose nature depends on the definition of the code. For bit flip codes, a logical \texttt{0} is encoded on the code qubits as $|0\rangle^{\otimes d}$ and $1$ is $|1\rangle^{\otimes d}$. These will detect bit flips during the idle time, i.e. errors composed of the Paulis $X$ and $Y$. For phase flip codes the codewords are $|+\rangle^{\otimes d}$ and $|-\rangle^{\otimes d}$, and it is the phase flips $Y$ and $Z$ that are detected. A fully general single qubit error could be composed of $X$, $Y$ and $Z$, but no single repetition code can detect a full set of errors in a single instance. However, by performing experiments with both bit flip and phase flip codes we can find a full set of error probabilities.

Given the above considerations, we now have a minimal set of experiments to probe the idling errors of a single code qubit during measurement of its auxiliaries. We will use $d=3$ repetition codes for $T=2$ syndrome measurement rounds, with the selected qubit in the centre. We extract the probability for an error on the central qubit during the first syndrome measurement round. This is done for both the bit flip encoding and phase flip encoding to get probabilities for both bit and phase flips.

\subsection{Constructing the circuits}

\begin{figure}[htbp]
\begin{center}
\includegraphics[width=\columnwidth]{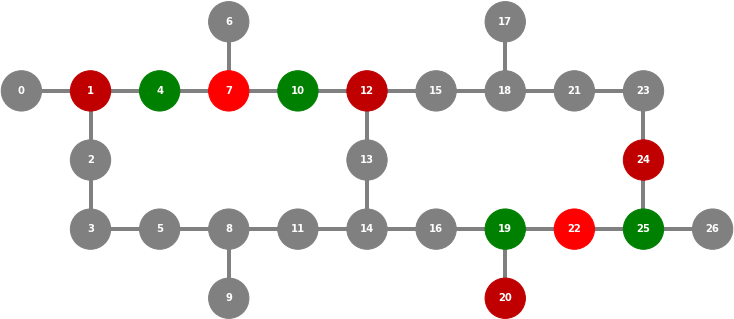}
\caption{Two lines of five qubits are shown on the coupling map of IBM Quantum \emph{Falcon} devices. For each, code qubits are shown in red and auxiliary qubits in green. One of these lines is the unique one required to benchmark qubit 7 on this device. The other is one of several options for qubit 22.}
\label{lines}
\end{center}
\end{figure}

For any given device, we must first determine which qubits can be analyzed in the manner described above. This will depend on the coupling graph, in which the qubits are vertices and edges correspond to pairs of qubits for which the two qubit \texttt{cx} gate can be applied. A qubit can only serve as the central qubit of a $d=3$ code if there exists a line of five qubits on this graph, centered on the chosen qubit. Examples of such lines are shown in Fig. \ref{lines}.

For some qubits, there is only one possible line of five qubits for which it sits at the center. For others, there are multiple options. For example, Fig. \ref{lines} shows the line $20-19-22-25-24$, which can be used to benchmark qubit 22. However, this could equally begin at 16 rather than 20, and end at 26 rather than 24.

In such cases where there is more than one option, more information is required. Our experiments are performed on IBM Quantum hardware via Qiskit. As such, there is an array of benchmarking data available regarding each device, which is updated during regular calibrations. In particular we will make use of the measured $T_1$ and $T_2$ times for each qubit, and the error probability of each \texttt{cx} gate.

The \texttt{cx} gate errors are what we use to determine the optimal line. Specifically, two quantities are determined for the \texttt{cx} gates within each line:
\begin{itemize}
\item The maximum error rate for \texttt{cx} gates that involve the central qubit;
\item The maximum error rate for \texttt{cx} gates within the line as a whole.
\end{itemize}
The chosen line is that which first minimizes the former, and then minimizes the latter. All lines involving a \texttt{cx} with an error rate of over $0.5$ are discarded, since these typically signal a poorly calibrated gate.

The `leaf qubits' around the edge, which couple only to a single neighbour, cannot serve as the central qubit in any such code. We therefore cannot study these in the same manner. Also, if the quality of some \texttt{cx} gates are sufficiently low, we may also be blocked from benchmarking further qubits.

\begin{figure}[htbp]
\begin{center}
\includegraphics[width=\columnwidth]{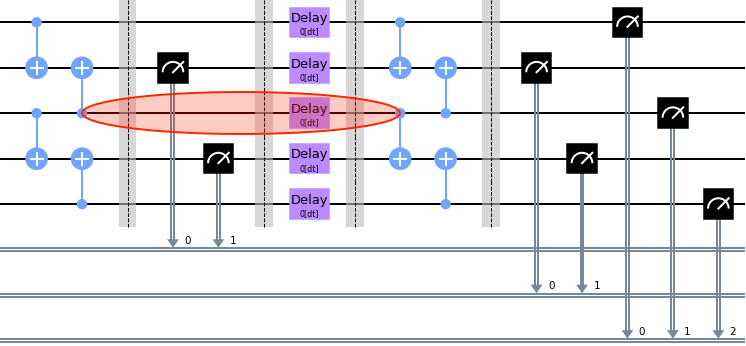}
\caption{The circuit used for codes with the bit flip encoding and a logical \texttt{0}. The position of the \texttt{delay} gates show where delays can be added to increase the effects of relaxation noise and dephasing. The errors measured by our experiments are those corresponding to the highlighted part of the circuit.}
\label{circuit}
\end{center}
\end{figure}

An example of a circuit implemented along the line is shown in Fig. \ref{circuit}. Such circuits contain multiple points at which qubits are idle, such as while waiting for a neighbour to complete an entangling gate or measurement. To increase performance, we can insert dynamical decoupling sequences during all idle times. When doing so, we insert a specific CPMG sequence~\cite{carr:1954,meiboom:1958}: a pair of \texttt{x} gates.

\section{Comparing results with with $T_1$ and $T_2$}

Before blindly accepting the error probabilities calculated from syndrome measurement results, it is important to determine if they produce results in line with expectations. For this we can greatly increase the amount of idling error by increasing the idling time. This is done by simply introducing a delay after each set of measurements. For a sufficiently long delay time, the errors due to idling will be greatly dominant over those from imperfections in the two qubit gates. It should therefore be possible to estimate the error probabilities from standard benchmarking data. Specifically: the $T_1$ and $T_2$ times.

A useful feature of the repetition code is that the code qubits are in a product state. It is therefore simple to determine the expected errors from $T_1$ and $T_2$. Specifically, for a total idle time of $t$ for the code qubits, we can derive the following properties of $P_{\rm 0 \rightarrow 1}$ (the probability that a $|0\rangle$ will flip to $|1\rangle$), $P_{\rm 1 \rightarrow 0}$ (vice-versa), and $P_{+\leftrightarrow -}$ (the probability of flipping between $|+\rangle$ and $|-\rangle$).

\begin{equation}
\begin{split}
P_{0 \rightarrow 1}(t) + P_{1 \rightarrow 0}(t) &= 1 - e^{-t/T_1}, \\
\frac{ P_{0 \rightarrow 1}(t) }{ P_{1 \rightarrow 0}(t) } &= \frac{1-p_0}{p_0}, \\
P_{+ \leftrightarrow -}(t) &= \frac{1 - e^{-t/T_2}}{2}.
\end{split}
\end{equation}

Here $p_0$ is the probability that the qubit will be found in the $|0\rangle$ state when the it reaches equilibrium. A well made qubit typically decays to a good approximation of the $|0\rangle$ state, and so $p_0 \approx 1$. This implies $P_{0 \rightarrow 1} \approx 0$ and $P_{1 \rightarrow 0} \approx 1 - e^{-t/T_1}$.

Before applying these probabilities to our states, we must account for the possibility of dynamical decoupling. Though methods for this vary, typically they will cause a qubit during a delay to spend half its time in a flipped state. For the bit flip encoding, this means that $t/2$ will be spend as both $|0\rangle$ and $|1\rangle$ regardless of the logical value. The relevant probability of a bit flip for either logical value during a delay is then,

\begin{equation} \nonumber
P_{0 \leftrightarrow 1}(t)  \approx P_{0 \rightarrow 1}(t/2) + P_{1 \rightarrow 0}(t/2) = 1 - e^{-(t/2)/T_1}.
\end{equation}

For phase flips, the use of some form of dynamical decoupling is already assumed when using the $T_2$ time. The probability $P_{+ \leftrightarrow -}(t)$ as stated above therefore applies in this case. For a circuit without dynamical decoupling, dephasing would instead be characterized by a shorter timescale $T_2^*$. This accounts for additional effects that are echoed out by the dynamical decoupling.

When adding a long delay, we will use $t=T/8$ where $T$ is the relevant timescale for the code being run. Specifically, $T=T_1$ For the bit flip encoding and $T=T_2$ for the phase flip encoding. With this delay time, we find,

\begin{equation}
P_{1 \rightarrow 0} \approx 11.8 \%, \,\,\,\, P_{0 \leftrightarrow 1} \approx 6.1 \%, \,\,\,\, P_{+ \leftrightarrow -}  \approx 5.9 \%.
\end{equation}

Without dynamical decoupling, the value of $P_{+ \leftrightarrow -}$ cannot be predicted from $T_2$ alone. However, we can expect that it will be comparatively high.

Note that these numbers are for the probabilities of a flip during the delay alone. There is also the probability of a flip during the circuit before and after the delay. Indeed this latter probability is what we otherwise wish to measure. However, we can assume that it will be relatively small in comparison. We therefore do not expect the measured flip probabilities to exactly align with the values above. Instead these values provide a guide to what kind of results we can regard as reasonable.

Codes were run on \texttt{ibm\_hanoi} to compare with these values. Both bit and phase flip encodings were run, both with and without dynamical decoupling. This was done for all the qubits on the device that could be placed at the center of a five qubit line. The median values found for the above probabilities were,

\begin{equation}
\hat{P}_{1 \rightarrow 0} = 16.0 \%, \,\,\,\, \hat{P}_{0 \leftrightarrow 1} = 9.6 \%, \,\,\,\, \hat{P}_{+ \leftrightarrow -}  = 25.0 \%.
\end{equation}

The first of these was determined from a run without dynamical decoupling, and the latter two with dynamical decoupling.

\begin{figure}[htbp]
\includegraphics[width=\columnwidth]{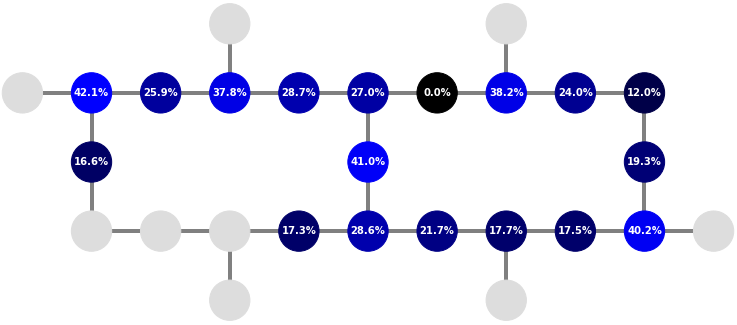}
\caption{The error probabilities ${P}_{+ \leftrightarrow -}$ for an additional delay time $t_{delay} = T/8$ and dynamical decoupling applied all qubits. The brightness of the qubits is proportional to the values.}
\label{long_results}
\end{figure}

The values for $\hat{P}_{1 \rightarrow 0}  =  16.0 \%$ and $\hat{P}_{1 \leftrightarrow 0} = 9.6\%$ are obviously larger than the guide values, but are nevertheless well within a factor of two. We therefore regard them as being within expectations. However, the value of $\hat{P}_{+ \leftrightarrow -}$ is significantly worse than the guide value of $5.9 \%$. The full set of probabilities for each qubit in this case is shown in Fig. \ref{long_results}, which shows that several qubits have even more extreme examples.

One possible mechanism that could cause additional dephasing is that described in~\cite{govia:2022}, where it was shown that decay events can cause phase noise on neighbouring qubits. In our case the neighbours of all code qubits are auxiliary qubits. The circuit only flips these to the $|1\rangle$ state when an error is detected. They should therefore mostly be in the $|0\rangle$ state, which would make them immune from relaxation errors. Since they should always be in either the $|0\rangle$ state or $|1\rangle$ state during delays, they are also immune to dephasing. The application of dynamical decoupling to these qubits will therefore have only negative effects. The biggest being the increased probability of relaxation errors when holding the qubit in the $|1\rangle$ state for long times, and therefore increased phase noise on the neighbouring code qubits. Our use of dynamical decoupling on all qubits may therefore have in fact caused this much increased dephasing.

\begin{figure}[htbp]
\includegraphics[width=\columnwidth]{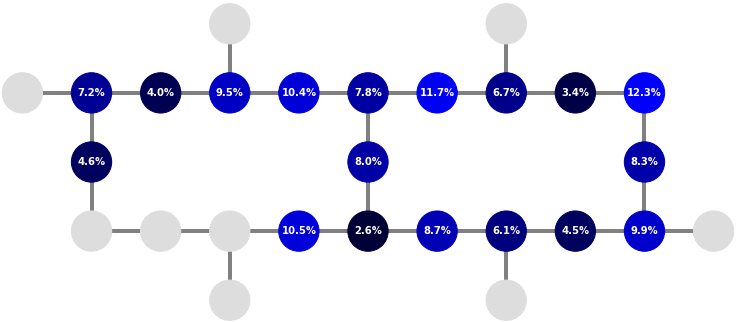}
\caption{The error probabilities ${P}_{+ \leftrightarrow -}$ for an additional delay time $t_{delay} = T/8$  and dynamical decoupling applied only to code qubits.}
\label{long_results_alt}
\end{figure}

Motivated by this, the experiments were run with dynamical decoupling only on code qubits. In this case we found $\hat{P}_{+ \leftrightarrow -}  = 7.95 \%$, which is in line with expectations. The full set of probabilities for each qubit in this case is shown in Fig. \ref{long_results_alt}.

Though implementing dynamical decoupling on the auxiliary qubits was a mistake, it was one that proved instructive. The complex noise triggered by a seemingly innocuous over-application of dynamical decoupling teaches an important lesson: we cannot assume that errors will always be described by a naive combination of $T_1$, $T_2$ and probabilities from randomized benchmarking. Measuring syndrome-derived error rates is therefore a means by which we can find and highlight these mysteries.

\section{Results from benchmarking devices}

\begin{figure*}
\centering
\begin{minipage}[b]{0.95\columnwidth}
\texttt{ibm\_hanoi}
\includegraphics[width=\w\columnwidth]{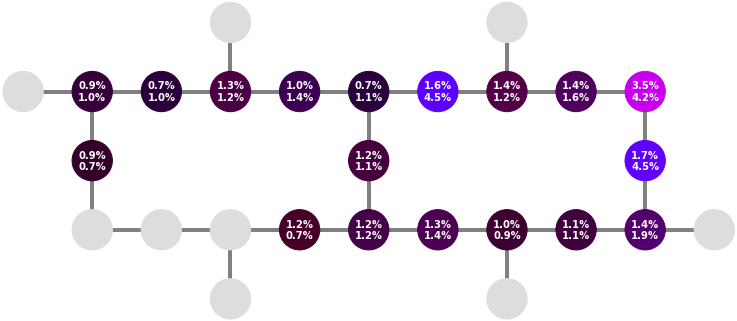}
\texttt{ibm\_auckland}
\includegraphics[width=\w\columnwidth]{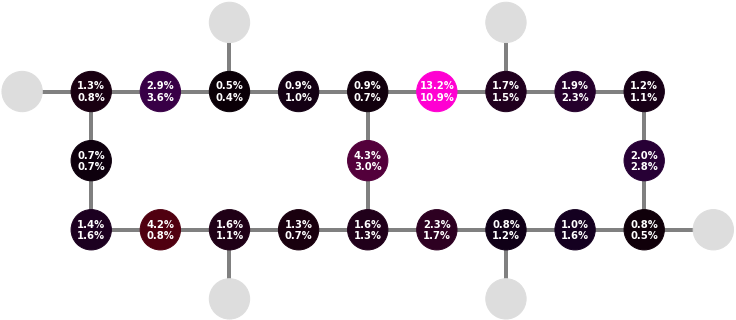}
\texttt{ibm\_cairo}
\includegraphics[width=\w\columnwidth]{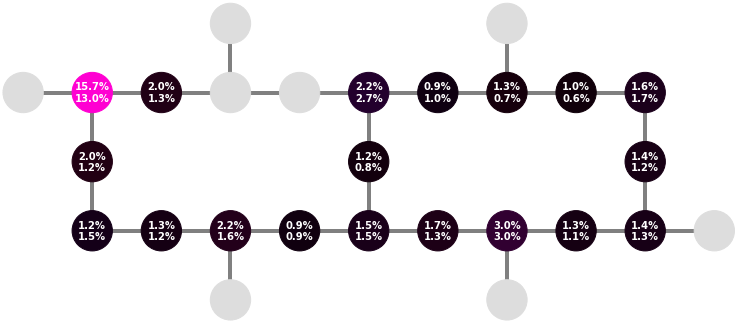}
\texttt{ibmq\_montreal}
\includegraphics[width=\w\columnwidth]{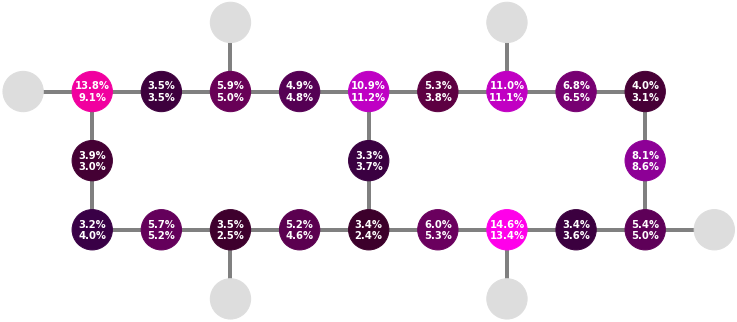}
\texttt{ibmq\_mumbai}
\includegraphics[width=\w\columnwidth]{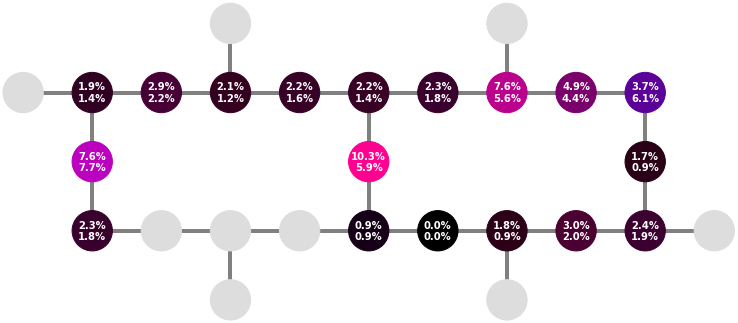}
\caption{This figure shows the probabilities ${P}_{0 \leftrightarrow 1}$ and ${P}_{+ \leftrightarrow -}$ measured on the qubits of a range of devices. The values are represented both in text (${P}_{0 \leftrightarrow 1}$ listed first and ${P}_{+ \leftrightarrow -}$ listed second), and in the colour of the qubit. For the latter, ${P}_{0 \leftrightarrow 1}$ and ${P}_{+ \leftrightarrow -}$ correspond to the brightness of the red and blue channels, respectively, with the maximum probability on the device used for the maximum brightness.}
\label{short_results}
\end{minipage}\qquad
\begin{minipage}[b]{0.95\columnwidth}
\texttt{ibm\_hanoi}
\includegraphics[width=\w\columnwidth]{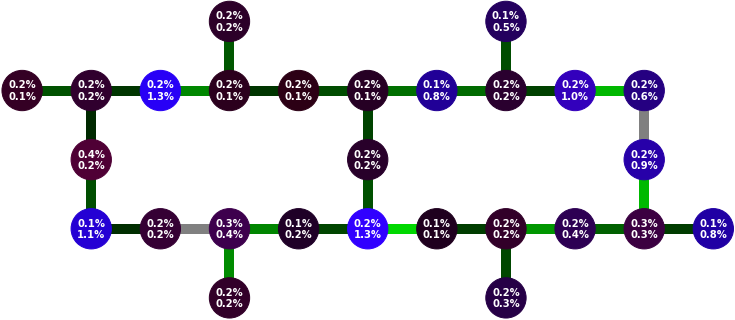}
\texttt{ibm\_auckland}
\includegraphics[width=\w\columnwidth]{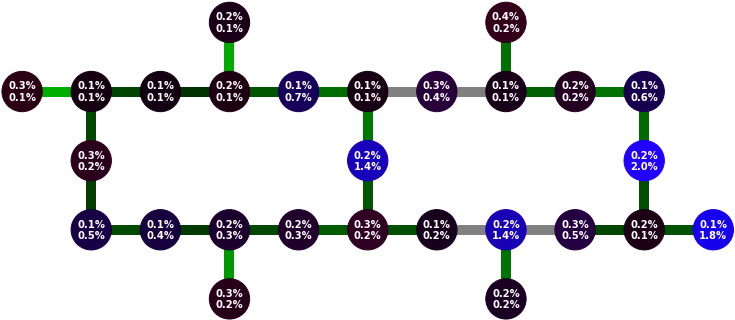}
\texttt{ibm\_cairo}
\includegraphics[width=\w\columnwidth]{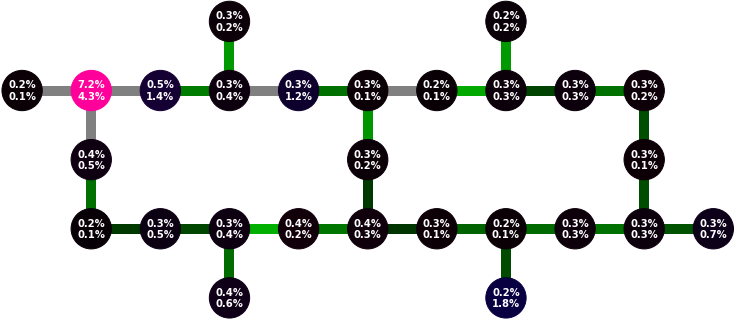}
\texttt{ibmq\_montreal}
\includegraphics[width=\w\columnwidth]{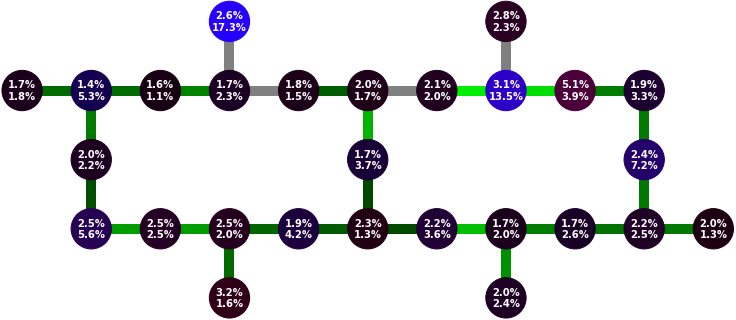}
\texttt{ibmq\_mumbai}
\includegraphics[width=\w\columnwidth]{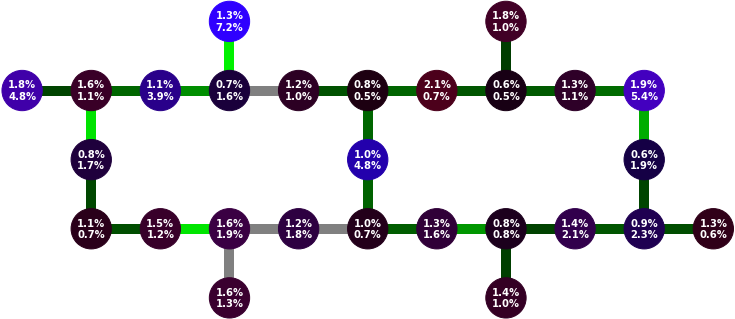}
\caption{The figure shows values for the probabilities ${P}_{0 \leftrightarrow 1}$ and ${P}_{+ \leftrightarrow -}$, calculated according to the $T_1$ and $T_2$ times of the qubits and the readout time on the device. The values are represented as in Fig. \ref{short_results}. The error rate for \texttt{cx} gates is represented by the colour of the corresponding link, with bright green corresponding to a $2\%$ error rate in all cases. Any \texttt{cx} with a higher error rate is shown in grey.
\\}
\label{ts_cxs}
\end{minipage}
\end{figure*}

In the case of dynamical decoupling applied to the code qubits alone, we have now found that the syndrome-derived error rates are largely within expectations. Given that we can expect to find sensible results, we can now proceed to our main goal: determine the idling errors of a single code qubit during measurement of its auxiliaries. For this we run the same circuits as above, but without the additional delay time. The results are shown in Fig. \ref{short_results}. The expected contributions from the $T_1$, $T_2$ and  \texttt{cx} errors, obtained from each device via Qiskit, are shown in Fig. \ref{ts_cxs}.

With these results, we can summarize the status of the worst and best qubits on each device (from the perspective of the syndrome-derived error rates). The numbering used for the qubits here is that seen in Fig. \ref{lines}.
\begin{itemize}
\item \texttt{ibm\_hanoi}: The worst qubit is 23, which has relatively long $T_1$ and $T_2$, but one of the worst \texttt{cx} gates; One of the best qubits is 4, which has a relatively long $T_1$ and typical \texttt{cx} gates for the device.
\item \texttt{ibm\_auckland}: The worst qubit is 15, which has a short $T_1$, a typical $T_2$ and two of the worst \texttt{cx} gates. The best qubit is 7, which has a relatively typical $T_1$, a long $T_2$, and good \texttt{cx} gates.
\item \texttt{ibm\_cairo}: The worst qubit is 1, which has short $T_1$ and $T_2$ and two of the worst \texttt{cx} gates. One of the best qubits is 15, with relatively long $T_1$ and $T_2$, but also has one of the worst \texttt{cx} gates.
\item \texttt{ibmq\_montreal}: The worst qubits are 1 and 19, despite having relatively typical $T_1$, $T_2$ and \texttt{cx} gates. The best qubit is 14, which has good $T_1$, $T_2$ and \texttt{cx} gates.
\item \texttt{ibmq\_mumbai}: The worst qubit is 13, which has among the worst $T_1$ ad $T_2$ times. The best is 16 which has relatively typical  $T_1$ and $T_2$, and relatively good \texttt{cx} gates.
\end{itemize}
Overall the results suggest that qubits with long $T_1$ and $T_2$ and low errors for \texttt{cx} gates will do well, and those for which either or both are very bad will do poorly. However, exceptions do occur, with the results from \texttt{ibmq\_montreal} in particular showing very poor behaviour from qubits whose $T_1$, $T_2$ appear typical for the device. The results therefore suggest an optimistic but cautionary outlook: the $T_1$, $T_2$ will give a good guide to the performance of a qubit within a quantum error correcting code, but direct measurement of the syndrome-derived error rates is required to ensure all is as it should be.

\section{Conclusions}

The implementation of small quantum error correcting codes is already well underway. Those experiments bring a wealth of syndrome information, which could provide interesting insights into the qubits of the devices. If the reader takes anything away from this paper, it should be that calculating and presenting these syndrome derived error rates should become standard practice. This would allow devices to be compared in a manner that lies between the extremes of randomized benchmarking and quantum volume~\cite{cross:19}, and which directly benchmarks performance towards large-scale quantum error correction.

In this work we investigated a minimal implementation of such a study, assessing qubits across a device with tailored $d=3$, $T=2$ repetition codes. Continuing to calculate these probabilities will provide users of quantum devices with a novel set of benchmarks for each device, giving a sense of how qubits will fare within complex circuits with mid-circuit measurement.

However, even these calculations are small and simple in comparison with the large-scale error correction of fault-tolerance. This work is therefore just the first step towards the continual assessment of quantum hardware through syndrome-derived error rates.

The source code and data for all results in this paper will be made available at~\cite{data4papers}.

\section{Acknowledgements}
The author would like to thank Daniel Miller for useful conversations.

Research was sponsored by the Army Research Office and was accomplished under Grant Number W911NF-21-1-0002. The views and conclusions contained in this document are those of the authors and should not be interpreted as representing the official policies, either expressed or implied, of the Army Research Office or the U.S. Government. The U.S. Government is authorized to reproduce and distribute reprints for Government purposes notwithstanding any copyright notation herein.

Part of the software used was created as part of a project supported by the NCCR SPIN, a National Centre of Competence in Research, funded by the Swiss National Science Foundation (grant number 51NF40-180604).

\bibliography{references}

\end{document}